# Maximum Lyapunov exponent revisited: Long-term attractor divergence of gait dynamics is highly sensitive to the noise structure of stride intervals.

PHILIPPE TERRIER & FABIENNE REYNARD
*Clinique romande de réadaptation, Sion, Switzerland.*

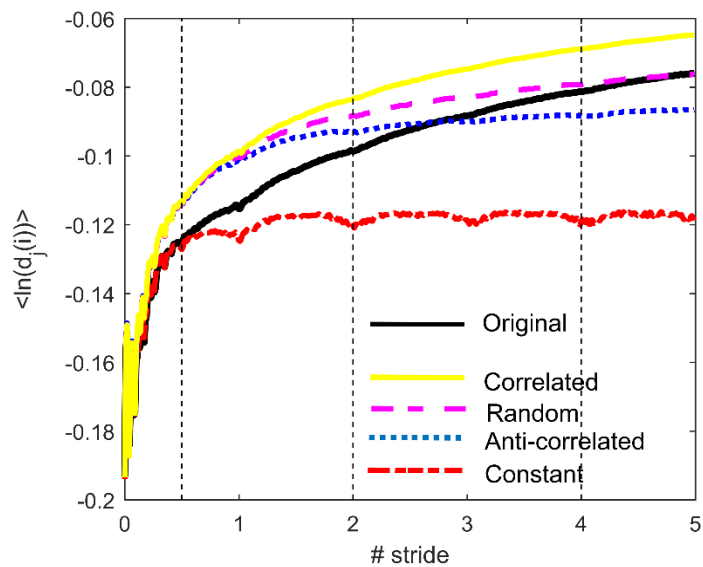

**Fig 2.** Divergence curves. A 5-dimensional attractor was built through the delay-embedding theorem from the trunk acceleration signal. Divergence between neighboring trajectories in the attractor is shown. For each noise type, 109 curves from 69 individuals were aggregated. Original stride intervals were changed to four different noise structures (hybrid signals). X-axis = time $i$ normalized by stride. Y-axis = the logarithm of the $i^{th}$ Euclidian distance $d$ downstream of the $j^{th}$ pair of the nearest neighbors in the attractor, averaged over all the pairs: $<\ln[d_j(i)]>$.

**Note:** This is a preprint of an article that was submitted for publication in February 2018.

**Abstract**

*Background:* The local dynamic stability method (maximum Lyapunov exponent) can assess gait stability. Two variants of the method exist: the short-term divergence exponent (DE), and the long-term DE. Only the short-term DE can predict fall risk. However, the significance of long-term DE has been unclear so far. Some studies have suggested that the complex, fractal-like structure of fluctuations among consecutive strides correlates with long-term DE. The aim, therefore, was to assess whether the long-term DE is a gait complexity index.

*Methods:* The study reanalyzed a dataset of trunk accelerations from 100 healthy adults walking at preferred speed on a treadmill for 10 minutes. By interpolation, the stride intervals were modified within the acceleration signals for the purpose of conserving the original shape of the signal, while imposing a known stride-to-stride fluctuation structure. Four types of hybrid signals with different noise structures were built: constant, anti-correlated, random, and correlated (fractal). Short- and long-term DEs were then computed.

*Results:* The results show that long-term DEs, but not short-term DEs, are sensitive to the noise structure of stride intervals. For example, it was that observed that random hybrid signals exhibited significantly lower long-term DEs than hybrid correlated signals did (0.100 vs 0.144, i.e. a 44% difference). Long-term DEs from constant hybrid signals were close to zero (0.006). Conversely, short-term DEs of anti-correlated, random, and correlated hybrid signals were closely grouped (2.49, 2.50, and 2.51).

*Conclusions:* The short-term DE and the long-term DE, although they are both computed from divergence curves, should not be interpreted in a similar way. The long-term DE is very likely an index of gait complexity, which may be associated with gait automaticity or cautiousness. Consequently, to better differentiate between short- and long-term DEs, the use of the term attractor complexity index (ACI) is proposed for the latter.

1. **Introduction**

Analysis of the nonlinear variability of human locomotion has attracted growing interest over the past decade [1]. This approach postulates that decoding nonlinear dependence among consecutive gait cycles (strides) can help to better understand gait control. A popular nonlinear method is the local dynamic stability (LDS) of the gait [2-5]. LDS is derived from the maximum Lyapunov exponent, which is used to highlight the deterministic chaos in nonlinear systems. Gait LDS has been proven particularly useful for detecting patients at risk of falling [6].

The majority of LDS studies use the Rosenstein's algorithm that computes the distance between trajectories of an attractor reflecting the gait dynamics [3, 4]. A logarithmic divergence curve is then built to assess the exponential divergence rate—the divergent exponent (DE)—by means of linear fitting over a given range. Two ranges have been proposed: a short-term range over 0–1 stride or step (the short-term DE), and a long-term range over 6–10 strides (the long-term DE) [4]. Puzzling results have been found when these two LDS indexes are used together to assess fall risk: both indexes most often vary in opposite directions [7, 8]. Further theoretical and experimental studies have shown that only the short-term DE is a valid gait stability measure [2, 9, 10]. However, it is not excluded that the long-term DE is associated with other gait features given its responsiveness to various conditions [11-13].

Another approach for studying nonlinear gait variability is the analysis of the noise structure of stride-to-stride fluctuations. Basic gait parameters, such as stride interval, stride length and stride speed, fluctuate among strides within a narrow range of 2%–4% [14]. It has been shown that these fluctuations are not random, but exhibit long-range correlations and a scale-free, fractal-like pattern [14-16]. This particular noise structure is observed in many different physiological signals, and is considered a hallmark of the complexity of living-beings [17]. Interestingly, this fractal structure can be altered when external cues are used to intentionally drive the steps, such as synchronizing gait to a metronome, or by following marks on the floor [16].

In 2009, Jordan et al. [18] analyzed both gait stability and complexity in treadmill walking and running. They observed a strong correlation ($r = 0.80$) between a measure of gait complexity (the scaling exponent of stride intervals) and the long-term DE. In 2012, Sejdic et al. [12] assessed the noise structure of stride intervals as well as the LDS (short-term and long-term DEs) during normal walking with and without external cueing (metronome walking). The results showed that, with auditory cueing, the long range-correlations of stride intervals changed to anti-correlated patterns along with a substantial decrease of long-term DEs, but with no change of short-term DEs. Similarly, in 2013 [4], we analyzed the gait stability and complexity of treadmill walking, confirming that both long-term DE and the noise structure of stride intervals were similarly modified by external cueing. A significant correlation between scaling exponents and long-term DEs ($r = 0.57$) was also observed. In summary, the long-term DE seems more associated with the noise structure of stride intervals than with local stability and fall risk. Complex fluctuations that occur over dozens of consecutive strides seems to induce a delayed plateauing of the divergence curve, resulting in a higher long-term DE.

The current study's objective was to further explore whether the long-term DE should be interpreted as an index of gait complexity rather than an index of gait instability. To this end, stride intervals of natural gait acceleration signals were replaced with artificial time series exhibiting known noise structure. The hypothesis was that higher long-term DEs were associated with a more complex variability of stride-to-stride fluctuations. It was also assumed that short-term DEs were, in contrast, not sensitive to the noise structure of stride intervals.

## 2. Methods

### 2.1 Setting

A large, anonymized dataset of acceleration signals obtained from our previous studies was re-analyzed [19, 20]. In short, 100 healthy individuals aged between 20 and 69 years walked at preferred speed on a treadmill for five minutes in two sessions, separated by one week. A 3D accelerometer (Physilog, GaitUp, Lausanne, Switzerland) with a 200Hz sampling rate, attached to the sternum, recorded the trunk acceleration.

### 2.2 Data pre-processing

Each of the two-hundred acceleration signals was pre-processed using Matlab (R 2017a; The Mathworks, Inc., Natick, MA, USA). First, the vertical signal was selected and normalized to zero mean (i.e. removal of the constant gravity component). Based on the walking cadence assessed using spectrum analysis, 500 steps (250 strides) were extracted from the 5-minute signal, which was then resampled to a constant length of 25,000 samples. A custom peak-detection algorithm found the local maxima, which corresponded to heel strikes (Fig. 1A, B). One in two of these maxima delimited each stride and constituted the original time series of stride intervals. The standard deviation (SD) and the coefficient of variation (CV = SD / mean x 100) characterized the variability magnitude among the stride intervals. Finally, the detrended fluctuation analysis (DFA) determined the noise structure of the stride-interval time series. DFA can detect self-similarity, and hence correlation structure, in non-stationary times series [4, 21]. The slope of a line-fit in a log-log plot of scales vs fluctuations is the scaling exponent. The evenly spacing method [22] was used, with box sizes between 6 and N/2, i.e.

125. If the scaling exponent is smaller than 0.5, the noise is deemed anti-correlated. Random noise has a scaling exponent of about 0.5. Correlated noise exhibit a scaling exponent lying between 0.5 and 1.

*2.3 Signal selection*

Using a simple algorithm to look for local maxima may produce spurious stride intervals due to the sporadic presence of two close acceleration peaks of similar intensity at heel strike. This phenomenon is mainly related to idiosyncratic gait pattern. Therefore, we excluded the poorly configured signals that corresponded to at least one of these two criteria: 1) an average CV of stride intervals greater than 8%; or 2) a scaling exponent below 0.5. The interval time series of the included signals, therefore, had noise structure and magnitude similar to the commonly admitted values, i.e. a CV around 3%, and a scaling exponent around 0.7 [16].

*2.4 Artificial times series*

For each included acceleration signal, we built three computer-generated time series with the same length (i.e. 250), mean, and SD as the original stride-interval time series, but with different noise structures (i.e. anti-correlated, random and correlated structures) (Fig. 1C). The random time series were generated by the Matlab random number generator (*normrnd*). Correlated and anti-correlated time series were generated with an autoregressive fractionally integrated moving average (ARFIMA) noise simulator [23]. Based on the time-series theory introduced by Box & Jenkins [24], ARFIMA models can simulate processes with long-range correlations among consecutive samples [25]. DFA was applied to measure the actual scaling exponent of the artificial time series.

*2.5 Hybrid signals*

Each acceleration signal was combined with the corresponding artificial time series to form the hybrid signals. We sought to preserve the shape of the original signal, while altering the duration of each stride according to the artificial time series. To this end, each stride in the original signal was extended or contracted by adding or removing points by interpolation (Fig. 1D). The stride intervals in the original signal were replaced by the intervals in the artificial times series. We used the shape-preserving piecewise cubic interpolation algorithm (*pchip*) provided by the Matlab function *interp1*. As a result, three hybrid signals with identical shape, but different noise structures for stride intervals, were obtained. A fourth hybrid signal was also generated by equalizing the duration of each stride to the mean stride interval (constant signal).

*2.6	Attractor divergence curves and divergence exponents*

Divergence curves and DEs were computed following the habitual method applied in our lab [4]. Multi-dimensional attractors were constructed based on the delay embedding theorem. A global false nearest neighbors (GFNN) algorithm determined an attractor dimension of five common for all signals. Individualized time delays were assessed by the average mutual information (AMI) of each signal. Logarithmic divergence curves were built with the Rosenstein's algorithm. The time axis (x-axis) was normalized by stride intervals. The average curves are presented in Fig. 2. The exponential divergence rate was computed for two time-scales: across the span of 0–0.5 stride (short-term DE) and 2–4 strides (long-term DE).

*2.7 Statistics*

Notched boxplots were used to show the distribution of the data. Means and SDs were also computed. Because of the hierarchical nature of the data (signals nested into subjects), we applied linear mixed models to assess the difference between the signal types, using *R* and the *lme4* package [26]. Dummy variables were used to differentiate among the conditions (independent variables). Short-term DE and long-term DE were the dependent variables and the subjects were the random effect (random intercept).

Model accuracy was assessed with the $R^2$ method of Nakagawa & Schielzeth [27]. We used the *R* package *R2GLMM* [28]. The marginal $R^2$ was computed; that is, the proportion of variance explained by the fixed effects. Semi-partial $R^2$ for each signal type was also computed.

**3.	Results**

*3.1 Stride-interval time series*

The analyzed database contained 200 acceleration signals. Among them, 109 signals from 69 participants were judged of sufficient quality to be included in the analyses. The included time series of stride intervals exhibited a mean CV of 2.9% (SD = 1.6%) and a correlated structure with a scaling exponent of 0.71 (Table 1). DFAs of the artificial times series confirmed that the ARFIMA algorithm generated the expected noise structure (Table 1).

*3.2 Divergence curves*

The average divergence curves showed a strong dependence on noise structure (Fig. 2). The original signals, with the natural stride-to-stride fluctuation structure, gave a steep divergence curve that nevertheless did not reach a plateau after five strides. Conversely, the hybrid signals with constant stride intervals produced a quickly dampened curve with no additional divergence after two strides. The three other signal types gave rise to a higher absolute divergence, with, nonetheless, distinct divergence rates. The curve of the correlated signals appeared parallel to the original signal, while the other two were more dampen.

*3.3 Short-term divergence exponent*

The noise structure of the stride intervals had no relevant effect on short-term DE. On average, hybrid signals with constant stride intervals were 6% lower than the original signals (Table 1). Conversely, the three other signals were slightly higher than the original signals (15%). Although the mixed linear model revealed that those differences were significant (Table 2), no differences were observed between the anti-correlated, random, and correlated signals (coefficients: 0.311, 0.326, and 0.332, respectively). Furthermore, 96% of the variance remained unexplained by the model ($R^2$ = 0.04), which shows the weakness of the association between noise structure and short-term DE.

*3.3 Long-term divergence exponent*

The noise structure of the stride intervals had a strong effect on long-term DE. Figure 2 clearly shows that the divergence rate tended to increase with signal complexity. That is, signals with constant stride intervals exhibited a near-zero long-term DE (0.006), while the other three hybrid signals were ordered as anti-correlated (0.049), random (0.100), and correlated (0.144) (Table 1). Note that the correlated hybrid signals presented a long-term DE (0.144) comparable to that of the original signal (0.162), which also had a correlated noise structure. Statistical inference from the mixed linear model confirmed the strong association between long-term DE and noise structure (Table 2); the fixed effects (i.e. the signal types) explained 72% of the long-term DE variance.

4. **Discussion**

The results supported the hypothesis that long-term DE is responsive to the noise structure of stride intervals. Indeed, hybrid signals with identical shapes but modified stride intervals had long-term DEs that varied strongly according to the type of noise applied. Alternatively, short-term DEs varied within a narrower range (~20%) and were not sensitive to noise type.

Regarding the analytical method, we did not use the classical range of 6-10 strides to compute long-term DE. This range was empirically chosen in the first study that proposed utilizing the Rosenstein's algorithm to assess gait stability [5]. Given that divergence curves flatten earlier than six

strides—typically around two strides—, other ranges may be more appropriate. In a recent study, we showed that a 2-to-6-range can differentiate between healthy people and patients suffering from chronic pain of lower limbs [13]. In the present study, we applied a 2-to-4 range that seemed appropriate to highlight the noise structure. However, further investigations are needed to characterize the repeatability and responsiveness of different ranges.

The small effects of stride interval manipulation on short-term DE (Fig. 2 and Table 2) are likely due to the magnitude of the added noise. Suppressing stride interval noise (constant signal) lowered the short-term DEs by 5%, whereas adding noise of the same magnitude as the original signal increased them by 15%. Longer or shorter strides are likely correlated to a specific shape of acceleration within the strides. The modification of stride intervals likely alters this correlation, which in turn produces slightly different divergence rate compared to the original signal. Notably, noise structure had not impact. This further supports that noise magnitude, not structure, is the cause of the observed differences.

Unlike short-term DE, long-term DE varied strongly according to noise type. Anti-correlated signals had a long-term DE 34% as high as the correlated signals (0.049 vs 0.144, i.e. a 66% difference). This result is in line with previous studies that highlight a correlation between scaling exponents and long-term DEs [4, 18]. Furthermore, we showed [4] that, when individuals walk on a treadmill following or not following the pace of a metronome, their stride intervals switched from a correlated structure (scaling exponent = 0.80) to an uncorrelated one (0.27); in parallel, long-term DE decreased by 70%. These previous results [4, 12, 18] and those of the present study demonstrate that long-term DE is an index of stride interval complexity and not an index of gait stability.

Modelling studies based on computer simulations of perturbed walking highlight that long-term and short-term DEs represent different aspects of gait dynamics [9, 10]. Interestingly, the passive dynamic walker used by Su & Dingwell exhibits a quickly dampened divergence curve (Fig. 6 in [10]) similar to the curve of the constant signal (Fig. 2); that is, a curve that becomes flat after two strides. Su & Dingwell explained that the model "does not incorporate any of the neuromuscular control mechanisms that humans have" [10]. Indeed, it is likely that the model did not correctly simulate the complex inter-stride fluctuations that cause the high long-term DE observed in humans.

Could the long-term DE be utilized in experimental and clinical studies to reveal relevant gait features? From a methodological point of view, it can be readily computed along with the short-term DE, which is used to assess fall risk. In addition, it is likely that the computing of long-term DE requires fewer strides than DFA. In contrast to DFA that necessitates the detection of stride occurrences, the Rosenstein's algorithm exploits the full acceleration signal. As pointed out by Rival et al. [29], gait analysis methods that do not rely on step detections may be more robust in certain pathologies and conditions that induce atypical acceleration signals (e.g. shuffling, crouched and toe gaits).

From physiological and clinical points of view, situations that require a close control of stepping, such as metronome walking [16], induce a loss of stride fluctuation complexity. We argue therefore that the more the gait is automatic, the higher the scaling exponent of the stride intervals and the long-term DE. The hypothesis that associates fluctuation complexity and gait automaticity (or, inversely, gait cautiousness) is supported by several observational studies. Herman et al. [30] show that older people at risk of falling have a cautious gait pattern (shorter steps, slower speed, and widening of the base of support) and a lower scaling exponent of their stride intervals. Similarly, healthy individuals adopt a more cautious pattern (shorter steps) when they are exposed to a destabilizing environment, which also induces lower long-term DEs [8]. Although these findings reinforce the idea that long-term DE could be responsive to gait cautiousness or automaticity, further studies are needed to better define its physiological significance and clinical usefulness.

## 5. Conclusion

The present study's findings further support the idea that the short-term DE and the long-term DE, although they are both computed from attractor divergence curves, should not be interpreted in a similar manner. Accordingly, we propose a new term to better differentiate between them. Because long-term DE is an index of complexity computed from a multidimensional attractor, the term attractor complexity index (ACI) is thought to be appropriate. We hope that the empirical clarification of the difference between LDS and the ACI will lead to innovative studies of the nonlinear dynamics of human gait.

**Acknowledgement**

The authors wish to thank Dr. Olivier Dériaz for his administrative help and useful advice. The SUVA and the clinique romande de réadaptation were the main sponsors of the study through internal funding. A gift from the Loterie Romande also supported the study. The Institute for Research in Rehabilitation is funded by the State of Valais and the City of Sion. The study's sponsors are not implied in the study design; the collection, analyses, or interpretation of data; the writing of the manuscript; or the decision to submit the manuscript for publication.

**Conflict of interest statement**

None declared.

**Authors' contributions**

Funding acquisition: PT. Conceptualization: PT. Methodology: PT. Data curation: PT, FR. Investigation: FR. Formal analysis: PT. Writing—original draft: PT. Writing—review and editing: PT, FR. Visualization: PT. Supervision: PT, FR. Project administration: FR. Both authors read and approved the final version of the manuscript.

**Table 1** Descriptive statistics.

| Signals (N=109) | Scaling exponent Mean (SD) | | Short-term DE Mean (SD) | | Long-term DE Mean (SD) | |
|---|---|---|---|---|---|---|
| Original | 0.71 | (0.14) | 2.18 | (0.81) | 0.162 | (0.043) |
| Constant | - | - | 2.05 | (0.78) | 0.006 | (0.016) |
| Anti-correlated | 0.22 | (0.03) | 2.49 | (0.92) | 0.049 | (0.020) |
| Random | 0.48 | (0.06) | 2.50 | (0.91) | 0.100 | (0.032) |
| Correlated | 0.86 | (0.08) | 2.51 | (0.93) | 0.144 | (0.055) |

DE = divergence exponent and SD = standard deviation.

**Table 2** Inferential statistics. Multivariate linear mixed models with short-term DE and long-term DE as dependent variables, and dummy variables representing the signal types as independent variables.

| | Predictors | Coefficient Estimate | 95% CI Lower | 95% CI Upper | Marginal $R^2$ | 95% CI Lower | 95% CI Upper |
|---|---|---|---|---|---|---|---|
| **Short-term DE** | | | | | 0.04 | 0.02 | 0.09 |
| | (intercept) | **2.238** | 2.034 | 2.442 | | | |
| Signal types | (Original) | | | | | | |
| | Constant | **-0.125** | -0.233 | -0.016 | 0.01 | 0.00 | 0.04 |
| | Anti-correlated | **0.311** | 0.203 | 0.420 | 0.01 | 0.00 | 0.04 |
| | Random | **0.326** | 0.217 | 0.434 | 0.01 | 0.00 | 0.04 |
| | Correlated | **0.332** | 0.224 | 0.441 | 0.00 | 0.00 | 0.02 |
| **Long-term DE** | | | | | 0.72 | 0.68 | 0.75 |
| | (intercept) | **0.162** | 0.154 | 0.170 | | | |
| Signal types | (Original) | | | | | | |
| | Constant | **-0.157** | -0.164 | -0.149 | 0.65 | 0.61 | 0.68 |
| | Anti-correlated | **-0.114** | -0.121 | -0.106 | 0.49 | 0.44 | 0.54 |
| | Random | **-0.062** | -0.070 | -0.055 | 0.22 | 0.17 | 0.28 |
| | Correlated | **-0.018** | -0.026 | -0.011 | 0.02 | 0.01 | 0.05 |

DE= divergence exponent and CI= confidence interval.

**Figures**

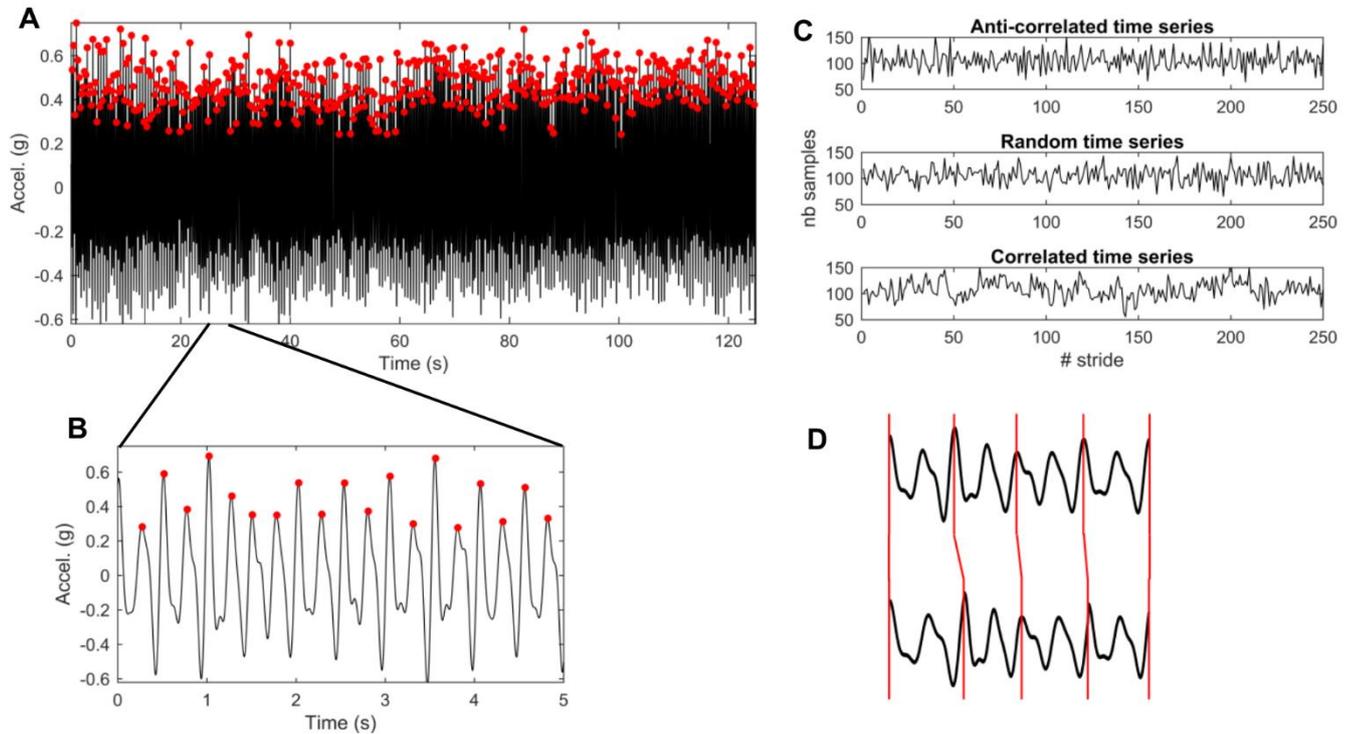

**Fig. 1.** Data processing. (A) Five minutes of vertical acceleration signal with steps detected by local maxima (red dots). (B) Magnification of the acceleration signal. (C) Examples of the artificial times series of 250 stride intervals generated with different noise structures. (D) Replacement of the original stride intervals through interpolation; the upper black trace is the original acceleration signal; the lower black trace is the interpolated signal; the vertical red lines indicate the original and transformed stride intervals.

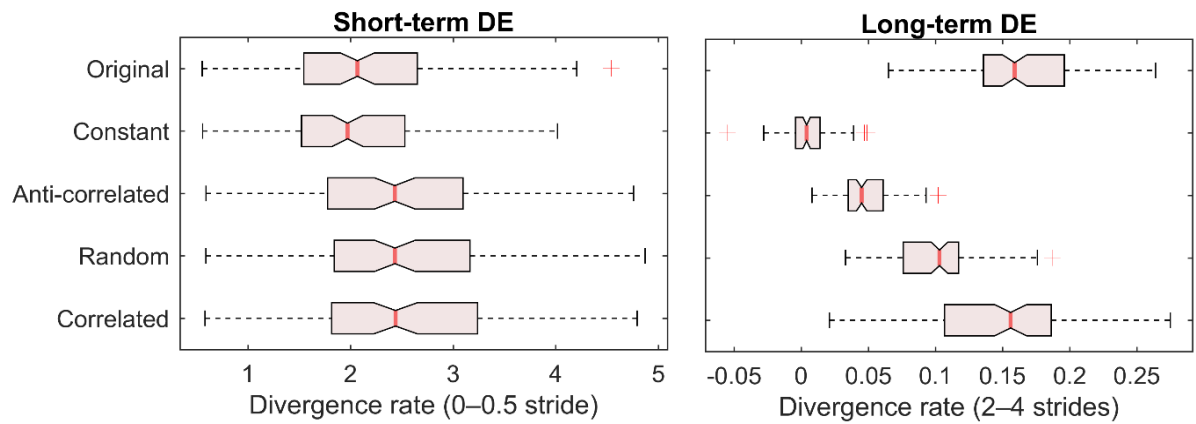

**Fig 3.** Distribution of the data. The boxplots show the quartiles, median, and extent of the data; the notches display the 95% confidence interval of the median; and the red crosses show the outliers.